\documentclass[fleqn,usenatbib]{mnras}

\usepackage{newtxtext,newtxmath}
\usepackage[T1]{fontenc}
\usepackage{ae,aecompl}
\usepackage{pdflscape}

\usepackage{graphicx}	% Including figure files
\usepackage{amsmath}	% Advanced maths commands
\usepackage{amssymb}	% Extra maths symbols

\title[The kinematics of Galactic disc white dwarfs in Gaia DR2]{The kinematics of Galactic disc white dwarfs in Gaia DR2}

\author[N. Rowell et al.]{
Nicholas Rowell,$^{1}$\thanks{E-mail: nr@roe.ac.uk (NR)}
Mukremin Kilic$^{2}$
\\
$^{1}$Institute for Astronomy, University of Edinburgh, Royal Observatory, Blackford Hill, Edinburgh EH9 3HJ, UK\\
$^{2}$Homer L. Dodge Department of Physics and Astronomy, University of Oklahoma, 440 W. Brooks St., Norman, OK, 73019, USA
}

\date{Accepted XXX. Received YYY; in original form ZZZ}

\pubyear{2018}

\begin{document}
\label{firstpage}
\pagerange{\pageref{firstpage}--\pageref{lastpage}}
\maketitle

\begin{abstract}
We present an analysis of the kinematics of Galactic disc white dwarf stars in the Solar neighbourhood using data
from Gaia Data Release 2. Selection of white dwarfs based on parallax provides the first large, kinematically unbiased
sample of Solar neighbourhood white dwarfs to date.
Various classical properties of the Solar neighbourhood kinematics have been detected for the first time
in the WD population.

The disc white dwarf population exhibits a correlation between absolute magnitude and mean age, which we exploit to
obtain an independent estimate of the Solar motion with respect to the Local Standard of Rest.
This is found to be $(U,V,W)_{\odot} = (9.5\pm1.2, 7.5\pm1.2, 8.2\pm1.2)$ kms$^{-1}$. The $UW$ components agree with 
studies based on main sequence stars, however the $V$ component differs and may be affected by systematics arising from
metallicity gradients in the disc.
The velocity ellipsoid is shown to vary strongly with magnitude, and exhibits a significant vertex deviation in the $UV$ plane
of around 15 degrees, due to the non-axisymmetric Galactic potential.

The results of this study provide an important input to proper motion surveys for white dwarfs, which require knowledge
of the velocity distribution in order to correct for missing low velocity stars that are culled from the sample to
reduce subdwarf contamination.
%
% A sentence or two about the velocity dispersion matrix
%
\end{abstract}


%%%%%%%%%%%%%%%%%%%%%%%%%%%%%%%%%%%%%%%%%%%%%%%%%%

%%%%%%%%%%%%%%%%% BODY OF PAPER %%%%%%%%%%%%%%%%%%

%
\section{Introduction}
The stars in the vicinity of the Sun move under the combined gravitational influence of all the other stars, gas and
dark matter in the Galaxy, and by analyzing their motions it is possible to infer aspects of the dynamical structure,
evolution and physical processes at work in the Milky Way as a whole. These include the kinematic heating of stars
through both secular processes and merger events, the local circular velocity and slope of the rotation curve, and
the shape of the gravitational potential.
% Asymmetry of the potential; asymmetric drift & age/velocity dispersion relation; Solar motion wrt LSR
%
Knowledge of the velocity distribution of stars is also relevant from a survey astronomy point of view, as it is a vital
input when quantifying the completeness of proper motion surveys (\citealp[e.g.][hereafter RH11]{rh2011}; \citealp{lam2019}).

This field has a long history, but went through something of a renaissance with the release of data from the Hipparcos mission 
\citep[e.g.][]{feast1997, db98, mignard2000, fvl2007, schonrich2010} which provided stringent constraints on various 
parameters of the Milky Way derived for the first time from absolute parallaxes.
In particular, \citet{db98} (hereafter DB98) used a sample of $16054$ main sequence (MS) stars to measure the Solar
motion with respect to the local standard of rest (LSR), an important quantity for transforming heliocentric velocities
to a Galactic frame. This classic technique exploits the correlation between colour and mean age that exists for
MS stars bluewards of the turn off of the old disk at $B-V\sim0.6$; the break in the trend of velocity dispersion
against colour that exists at this point is known as Parenago's discontinuity.
However, as explained by \citet{schonrich2010} the DB98 methodology is undermined by metallicity gradients within the
disc that lead to complex kinematic substructure in the colour--magnitude plane.
This field is now set to be truly revolutionized by the data from the Gaia mission \citep{gaia2016}, the successor to
Hipparcos, which ultimately aims to measure absolute parallaxes and proper motions at the microarcsecond level for more
than one billion stars in the Milky Way down to $G\approx20$.
The second data release \citep[DR2 -- see][]{gaia2018} was
published on 25th April 2018, and already contains over 1.3 billion sources with parallax accuracies of the order 700 $\mu$as
at the $G=20$ faint end. Much progress is still to be made at reducing systematic errors by improving various instrument
calibrations, but early studies demonstrate the enormous potential of the Gaia data.
These include \citet{bovy2017a}, who measured the Oort Constants that parameterise the local stellar streaming velocity,
obtaining the first significant estimates of the $C$ and $K$ terms. \citet{katz2018} used a subsample of 6.4 million FGK
stars with full 6D phase space coordinates to draw 3D maps of the stellar velocity distribution with unprecedented accuracy,
revealing rich substructure and complex streaming motions in the Galactic disk. On a larger scale, \citet{kawata2019} used
a sample of 218 Cepheids with Gaia proper motions to estimate several parameters of the Galactic rotation, including the
centrifugal velocity, the slope of the circular speed curve, the local velocity dispersion and the Sun's peculiar velocity.

%%%%%%%%%%%%%%%%%%%%%%%%%%%%%%%%%%%%%%%%%%%%%%%%%%%%%%%%%%%%%%

One avenue that has recently been opened by the Gaia DR2 is the
possibility of conducting a kinematic study using the white dwarf (WD) stars in the Solar neighbourhood.
Both Hipparcos and Gaia DR1 (TGAS) contain only very small numbers of WDs that cannot be used for a systematic study.
Instead, the largest quantifiably-complete catalogues of WDs to date have been derived using the Reduced Proper Motion (RPM) method applied
to ground based survey data. These include the Sloan Digital Sky Survey \citep{kilic2006, munn2017}, the SuperCOSMOS Sky Survey (RH11)
and PanSTARRS 1 \citep{lam2019}, each of which contain of order $\sim10000$ WDs. The problem with these catalogues, from a kinematic
point of view, is that the RPM method necessitates the use of a low tangential velocity threshold of around ~$30$kms$^{-1}$, below 
which stars are rejected in order to limit contamination from subdwarfs. This results in a catalogue that is suitable for luminosity
function studies but is inherently kinematically biased. Note that the SDSS spectroscopic WD catalogue \citep{kleinman2013,kepler2015}, which
contains around 30000 WDs in total, is biased by the complex selection function for fibre allocation.

The Gaia DR2 on the other hand contains several hundred thousand WDs \citep{2018arXiv180703315G} that can be reliably
identified based on parallax. The need to construct a robust sample within well characterized survey limits 
requires use of stricter selection criteria that reduces the useful sample size, but even so this is orders of magnitude
larger than any previous comparable WD survey.
Although in any magnitude-limited survey, such as that conducted by Gaia, WDs will exist in much smaller numbers than main
sequence stars, analysis of their kinematics is still of importance for the following reasons:
\begin{itemize}
    \item In a volume-limited sample such as the Solar neighbourhood the WDs make up a much more significant fraction of 
    the population, and any study that omits them can only be an incomplete picture.
    %This is especially true for studies that exploit the colour / age correlation that exists for MS stars lying
    %bluewards of Paranego's discontinuity, which concides with the turn off of the old disk.
    \item The location of a WD in colour-magnitude space is independent of the progenitor star metallicity, which may avoid
    certain systematics that arise in similar studies that use MS stars, as described in \citet{schonrich2010};
    briefly, metal-rich stars generally formed at $R<R_0$ and visit the Solar neighbourhood close to the apocentre 
    of their orbit where their azimuthal velocities lag behind the local circular speed. As metallicity displaces the MS, this
    introduces substructure in the colour-magnitude diagram that correlates with kinematics.
    \item For Solar neighbourhood WDs there exists a correlation between absolute magnitude and mean age, analogous to the
    colour/age correlation that exists for MS stars, which can be exploited to obtain
    an independent determination of the motion of the Sun with respect to the local standard of rest.
\end{itemize}
 
\begin{figure}
 \includegraphics[width=\columnwidth]{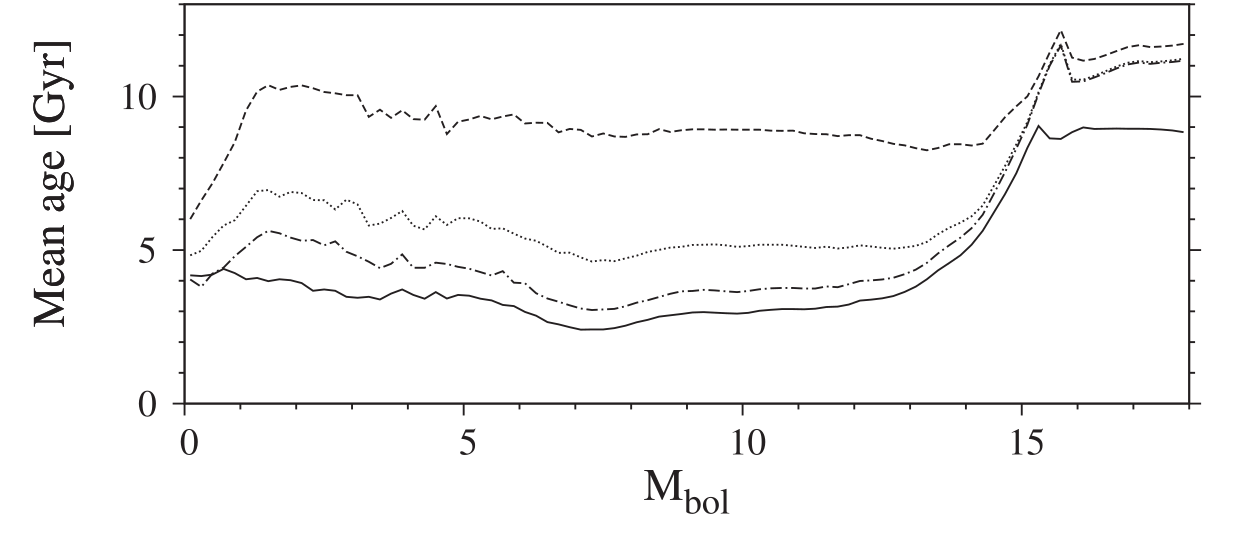}
 \caption{
 %\textbf{
 The mean total stellar age as a function of bolometric magnitude for a range of simulated white dwarf
 populations of different star formation rate $\psi(t)$. The solid and dot-dashed lines are for a constant $\psi$ with
 a total age $T_0=10$ and 13 Gyr respectively; the dashed line is for $\psi(t) \propto \exp (-t/\tau)$ with $\tau=3$ Gyr 
 and $T_0=13$ Gyr; the dotted line is for $\psi(t) \propto (1 + \exp ((t - t^{\prime})/\tau ))-1$ with $t^{\prime}=10$ Gyr,
 $\tau=3$ Gyr and $T_0=13$ Gyr. This figure is reproduced from \citet{Rowell2013}.
 %}
 }
 \label{fig:mean_age}
\end{figure}

The final point holds for any WD population that did not form in a single burst, and is due mainly to the monotonic cooling law
\citep{mestel1952}. This magnitude/age correlation exists for essentially all realistic
star formation histories, though the strength of the correlation varies and requires detailed modelling to quantify.
% Reponse to reviewer point 4)
%\textbf{
In figure \ref{fig:mean_age} (reproduced from \citet{Rowell2013}) we depict the mean total stellar age as a function of 
bolometric magnitude for a range of simulated white dwarf populations of different star formation history. The correlation
is strongest in the $M_{\mathrm{bol}}$ = 13 to 15 range and for the less burst-like star formation histories.
The Gaia G magnitude is expected to show a similar trend due to the extremely broad wavelength range leading to small
bolometric corrections of $\Delta M_{\mathrm{bol}} <= 0.09$ for both H and He atmosphere WDs over the $G$ = 13 to 15
range (P.~Bergeron, priv. communication).
%}
%

In this study we use Gaia DR2 to assemble a catalogue of 78,511 WDs in the solar neighbourhood.
This is the largest kinematically complete sample of WDs to date. We use this to investigate several classical features of the
Solar neighbourhood kinematics, including the asymmetric drift, local standard of rest and age-velocity dispersion relation.
Note that this study is independent of WD cooling models, being based on only direct observational quantities.
In section~\ref{sec:wd_selection} we outline the construction of our WD sample. Section~\ref{sec:methods} explains the methods used
to determine the kinematics, which are based on the proper motion deprojection method presented in DB98. We present our results in 
section~\ref{sec:res} with further discussion in section~\ref{sec:disc}, and draw conclusions in section~\ref{sec:conc}.
\section{White Dwarf selection}
\label{sec:wd_selection}
The sample of white dwarfs (WDs) used in this work was obtained from the Gaia data by selection in the Hertzsprung-Russell diagram (HRD).
We applied initial filters on the signal-to-noise of the photometry and astrometry to reject objects with relative parallax errors
greater than 20\% and errors on $G_{BP}$/$G_{RP}$ flux greater than 10\%. This reduces the scatter in the HRD and allows the absolute magnitude to
be determined based on simple inversion of the parallax. We also place cuts on the \emph{flux excess factor} to eliminate objects whose
photometry is compromised by blending issues, and we reject objects further than 250pc from the Sun for reasons explained in 
section~\ref{sec:methods}. This resulted in an initial sample of $3061480$ objects drawn from the Gaia archive; the ADQL query that 
performs this selection is presented in appendix~\ref{app:adql}.

\begin{figure}
 \includegraphics[width=\columnwidth]{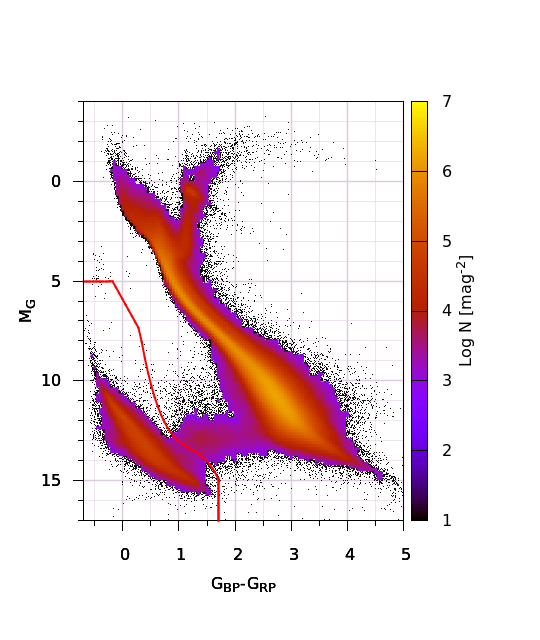}
 \caption{Hertzsprung-Russell diagram of $2644657$ well-measured disk stars within 250 pc. The red line marks the WD selection region
 in the Gaia bands defined by \citet{2018arXiv180703315G}, which contains $79225$ WDs. Sources have been filtered using the Renormalised
 Unit Weight Error (see text).}
 \label{fig:hrd}
\end{figure}

We applied additional astrometric quality criteria to reduce contamination from objects with poor astrometric solutions, by rejecting
objects with a Renormalised Unit Weight Error (RUWE; described in \citet{LL-124}) greater than $1.4$. This has the effect of removing
objects lying between the main sequence and white dwarf locus, which includes many partially resolved close pairs that the Gaia astrometric 
solution is not yet capable of handling properly. We also reject objects with tangential velocity greater than $150$kms$^{-1}$ as
these are likely not disk members. This reduces the initial sample to $2644657$ well-measured Solar neighbourhood disk stars. The
Hertzsprung-Russell diagram (HRD) of these objects is presented in figure~\ref{fig:hrd}.
%
%% HRD figure

\begin{figure}
 \includegraphics[width=\columnwidth]{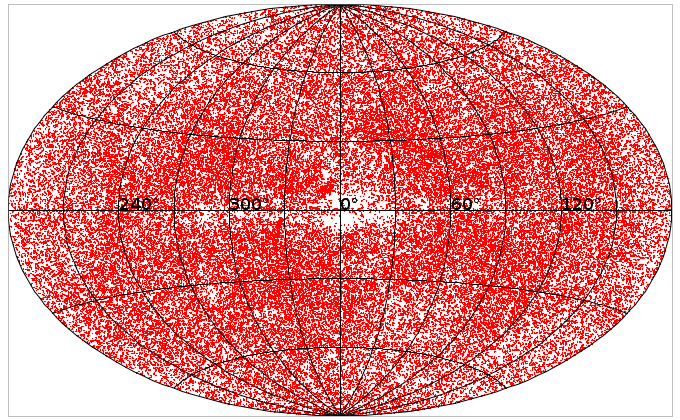}
 \caption{Sky distribution in Galactic coordinates of the $78511$ WDs that pass our selection criteria.
 %\textbf{
 The lower density of sources
 towards the Galactic centre is most likely due to problems that Gaia faces in crowded regions, as explained in the text.
 %}
 }
 \label{fig:sky_dist}
\end{figure}

%% WD selection
We select WDs from the lower left region of the $M_G$/$G_{BP}-G_{RP}$ plane using the same criteria as \citet{2018arXiv180703315G}. This region is
marked by a red boundary in figure~\ref{fig:hrd}, and contains $79225$ WDs.
There is also clearly a small amount of contamination from unresolved WD+dM systems lying in the region between the MS and WDs, which will
add some degree of noise to our results in the $M_G$ range $\approx13$--$15$. Although the astrometry for these objects is presumably reliable, having passed our selection criteria,
the photometry will be compromised. The fraction of contaminating sources as a function of $M_G$ can be estimated from the red
tail of the colour distribution; we estimate that the contamination is limited to less than a few percent at all magnitudes. We do not expect
this level of contamination to have a significant effect on our results.
We also restrict the analysis to WDs with absolute magnitudes in the range [10:15.5] for the following reasons. Brighter than $M_G=10$ there
are very few WDs and the correlation with age is nonexistent. Fainter than $M_G=15.5$, the WD population starts to become heavily contaminated
by thick disk and spheroid objects that have different kinematics, as well as low and high mass WDs that have not followed the same 
evolutionary  path as the rest of the population and so don't obey the same magnitude / mean age correlation.
This results in a clean sample of $78511$ WDs. The sky distribution of these is presented in figure~\ref{fig:sky_dist}.
%
% Response to reviewer point 1)
%\textbf{
Note that the lower density of stars towards the Galactic centre is not due to dust extinction, as our sample is relatively local.
Instead this is most likely caused by various issues that Gaia faces in crowded regions, in particular blending of objects in both the 
photometric and astrometric instruments, but also challenging cross-match and loss of faint stars due to onboard resource limitations.
%}
%
\section{Methods}
\label{sec:methods}
The WDs in Gaia DR2 have parallax and proper motion estimates but no radial velocities. Even the final Gaia catalogue will have
few radial velocities for WDs, due to their intrinsic faintness and the fact Gaia's radial velocity spectrograph is tuned
to the calcium triplet that does not appear in the spectra of the majority of WDs. Although the lack of radial velocities
means the 3D space motion of individual stars cannot be recovered, we can still calculate the moments of the 3D velocity
distribution by means of the proper motion deprojection technique introduced by DB98.
In this method, we first correct the proper motions of stars for the effects of Galactic rotation using equation~1 from DB98,
with values of the Oort constants $A=15.3$ and $B=-11.9$ taken from \citet{bovy2017a}.
The remaining tangential velocity of each star $\underline{p}$, which is estimated from the parallax
and corrected proper motion, is related to the 3D space motion $\underline{v}$ according to
\begin{equation}
\underline{p} = A \underline{v}
\end{equation}
where the projection matrix $A$ is defined by $\mathbb{I} - \underline{\hat{r}} \otimes \underline{\hat{r}}$, with $\underline{\hat{r}}$ 
the unit vector towards the star and $\otimes$ denotes the outer product.
Obviously, this equation cannot be inverted to obtain $\underline{v}$ and as such the matrix $A$ is singular. However, by averaging
over a large number of stars distributed widely across the whole sky, the mean matrix $\left<A\right>$ is non-singular and
the equation can be inverted to obtain the mean velocity $\left<\underline{v}\right>$ of the population relative to the Sun.
The residual velocities of stars can be used to estimate the scalar velocity dispersion, and an extension of this method allows
the full velocity dispersion tensor to be computed. We refer the reader to DB98 for further details.
As explained in \citet{mcmillan2009}, the method is strictly only applicable when the lines of sight to individual stars are uncorrelated
with the stars' velocities. This criteria holds for stars within a limited distance from the Sun, beyond which the velocity dispersion
varies with both radius and distance from the plane. For this reason we restrict the analysis to stars within 250pc. In any case,
the Gaia apparent magnitude limit of $G\mathbf{\approx}21$ means that fainter than around $M_G\mathbf{\approx}13.7$ all our stars are within 250pc anyway, and according to
\citet{Rowell2013} the correlation between magnitude and total age is very weak brighter than $M_G\mathbf{\approx}13$.

% Comparison with results from assuming zero radial velocity.
Note that this method is superior to simply assuming zero radial velocity and computing the sample mean and covariance, which
is done in some studies. The assumption of zero radial velocity causes the mean velocity to be underestimated by (on average)
one third. The velocity dispersion is also underestimated by up to one third due to a combination of the loss of the peculiar
velocity along the line of sight and the bias in the estimated mean.

We apply the kinematic analysis independently to bins of different G magnitude. We tried bins containing equal numbers of stars,
but this conceals important structure around the faint end of the range where there are relatively few stars. After some tests
we settled on bins of a fixed width of $\Delta M_G=0.125$.

\section{Results}
\label{sec:res}
We applied the methods described in section~\ref{sec:methods} to our sample of 78,511 WDs, independently on 44 bins of
$\Delta M_G=0.125$ between $M_G=10$ and $M_G=15.5$.
Table~\ref{tab:results} presents the mean velocity and velocity dispersion tensor for each magnitude bin.
In this section we highlight some selected results from these data.

\begin{table*}
\label{tab:results}
\begin{tabular}{|cc|ccc|cccccc|}
\hline
 Bin centre & N & \multicolumn{3}{|c|}{Mean velocity [kms$^{-1}$]} & \multicolumn{6}{|c|}{Velocity dispersion [km$^2$ s$^{-2}$]} \\
 $M_G$ [mag] & [stars] & V$_U$ & V$_V$ & V$_W$ & $\Sigma_{UU}$ & $\Sigma_{VV}$ & $\Sigma_{WW}$ & $\Sigma_{UV}$ & $\Sigma_{UW}$ & $\Sigma_{VW}$ \\
\hline
\hline
10.063 & 82 & -15.88 & -20.84 & -9.49 & 1846.13 & 627.04 & 227.49 & 562.22 & 461.50 & -207.45 \\
10.188 & 119 & -2.05 & -18.56 & -9.03 & 1331.90 & 685.98 & 259.71 & 225.83 & -341.21 & 146.84 \\
10.313 & 166 & -13.20 & -19.16 & -7.69 & 1381.77 & 550.74 & 140.47 & 263.70 & 129.24 & 321.92 \\
10.438 & 223 & -6.95 & -22.92 & -9.36 & 1407.75 & 626.93 & 250.58 & -1.85 & 22.80 & 70.62 \\
10.563 & 349 & -7.81 & -22.40 & -7.30 & 1171.79 & 522.92 & 417.56 & 15.16 & 27.70 & 82.71 \\
10.688 & 500 & -9.28 & -19.68 & -8.95 & 1118.97 & 471.78 & 266.60 & 350.77 & 73.70 & 66.07 \\
10.813 & 635 & -9.31 & -19.47 & -7.89 & 1248.14 & 592.03 & 295.02 & 165.80 & -40.07 & 18.47 \\
10.938 & 786 & -10.01 & -18.75 & -8.88 & 1257.76 & 498.45 & 341.81 & 299.97 & -112.53 & 12.59 \\
11.063 & 1048 & -8.19 & -20.91 & -9.15 & 1212.53 & 550.68 & 258.01 & 162.88 & 72.62 & 105.13 \\
11.188 & 1186 & -8.00 & -18.66 & -8.28 & 1328.73 & 508.88 & 308.12 & 232.68 & 85.83 & 116.67 \\
11.313 & 1390 & -9.55 & -20.28 & -8.00 & 1199.62 & 542.13 & 256.45 & 193.32 & -53.75 & -21.61 \\
11.438 & 1574 & -9.34 & -20.77 & -7.73 & 1036.90 & 516.40 & 281.83 & 118.39 & 87.60 & -19.37 \\
11.563 & 1690 & -10.25 & -20.02 & -8.25 & 1254.47 & 567.40 & 314.16 & 384.97 & 45.03 & -11.15 \\
11.688 & 2034 & -10.35 & -20.11 & -6.94 & 1245.43 & 553.62 & 313.30 & 195.89 & -0.53 & 16.19 \\
11.813 & 2102 & -9.89 & -20.07 & -8.43 & 1124.08 & 510.88 & 301.71 & 249.59 & 10.09 & 31.51 \\
11.938 & 2243 & -8.29 & -19.67 & -7.14 & 1196.07 & 510.06 & 275.11 & 162.78 & 6.94 & -19.39 \\
12.063 & 2224 & -9.97 & -19.05 & -7.28 & 1158.61 & 485.98 & 249.41 & 244.89 & -28.94 & 17.43 \\
12.188 & 2371 & -8.99 & -18.32 & -6.20 & 1177.38 & 544.20 & 248.29 & 266.90 & 27.85 & -7.90 \\
12.313 & 2500 & -8.34 & -18.70 & -8.00 & 1197.14 & 536.58 & 281.03 & 231.84 & 41.84 & 18.75 \\
12.438 & 2545 & -8.65 & -20.56 & -8.27 & 1147.77 & 581.53 & 339.41 & 268.62 & 25.24 & 43.65 \\
12.563 & 2801 & -8.89 & -19.74 & -7.07 & 1112.51 & 542.88 & 302.89 & 297.00 & 13.54 & -14.77 \\
12.688 & 2891 & -10.58 & -20.55 & -7.38 & 1151.91 & 553.23 & 314.96 & 257.92 & 40.65 & 12.79 \\
12.813 & 3133 & -10.27 & -20.95 & -8.48 & 1200.32 & 564.73 & 337.15 & 178.52 & 44.53 & 15.18 \\
12.938 & 3200 & -7.81 & -20.41 & -7.24 & 1248.96 & 533.83 & 302.04 & 198.85 & 16.37 & 75.86 \\
13.063 & 3287 & -8.95 & -21.30 & -6.39 & 1280.43 & 525.50 & 278.14 & 287.91 & 12.13 & 37.62 \\
13.188 & 3319 & -7.85 & -19.86 & -6.91 & 1246.33 & 536.79 & 307.74 & 289.40 & 71.29 & 36.38 \\
13.313 & 3221 & -8.86 & -21.29 & -8.07 & 1229.89 & 527.36 & 317.62 & 291.31 & -8.16 & 63.57 \\
13.438 & 3326 & -8.38 & -21.79 & -6.77 & 1315.21 & 607.53 & 336.40 & 206.81 & 68.29 & 47.08 \\
13.563 & 3218 & -7.87 & -21.90 & -7.41 & 1289.66 & 645.94 & 362.45 & 234.84 & 70.57 & 63.60 \\
13.688 & 3142 & -8.52 & -21.65 & -7.70 & 1449.37 & 526.23 & 344.46 & 254.54 & 39.18 & 44.32 \\
13.813 & 2953 & -9.72 & -22.54 & -6.61 & 1411.39 & 598.96 & 357.67 & 192.26 & -11.50 & 84.65 \\
13.938 & 2865 & -7.95 & -23.69 & -7.20 & 1472.20 & 616.75 & 357.78 & 255.87 & -34.56 & 51.51 \\
14.063 & 2675 & -10.31 & -24.21 & -7.64 & 1520.92 & 684.45 & 355.90 & 315.22 & 6.13 & 59.11 \\
14.188 & 2388 & -7.97 & -24.07 & -7.29 & 1543.36 & 653.46 & 441.34 & 278.35 & -66.81 & 68.07 \\
14.313 & 1965 & -11.16 & -24.42 & -7.82 & 1679.18 & 653.07 & 489.92 & 284.11 & 196.48 & 67.31 \\
14.438 & 1690 & -8.29 & -25.41 & -7.32 & 1621.61 & 762.39 & 404.23 & 195.83 & -55.39 & -23.96 \\
14.563 & 1513 & -10.15 & -25.80 & -6.88 & 1697.80 & 715.78 & 477.75 & 299.89 & -51.76 & -58.32 \\
14.688 & 1351 & -9.30 & -27.65 & -7.72 & 1849.70 & 684.15 & 544.75 & 150.33 & -94.65 & -52.84 \\
14.813 & 1225 & -7.39 & -28.81 & -7.73 & 2002.02 & 774.65 & 569.74 & 400.71 & 65.81 & 147.53 \\
14.938 & 931 & -8.34 & -28.43 & -5.51 & 1989.77 & 795.08 & 682.89 & 418.77 & -183.88 & 59.79 \\
15.063 & 687 & -8.77 & -32.44 & -5.42 & 2259.52 & 971.15 & 656.68 & 344.83 & 70.88 & 157.15 \\
15.188 & 475 & -14.64 & -33.43 & -6.47 & 2255.08 & 823.56 & 823.38 & 271.34 & 61.09 & 106.76 \\
15.313 & 300 & -14.50 & -41.76 & -0.98 & 2439.98 & 1153.29 & 935.68 & 43.40 & 14.91 & -32.18 \\
15.438 & 188 & -10.10 & -46.06 & -1.03 & 2078.84 & 1463.10 & 998.57 & -103.49 & 545.14 & -228.99 \\
\hline
\end{tabular}
\caption{For each magnitude bin, we present the mean stellar velocity relative to the Sun (columns 3--5) and the velocity
dispersion tensor (columns 6--11). Units are kms$^{-1}$ and km$^2$s$^{-2}$ respectively, and quantities are referred to the
Galactic coordinate frame with principal axes $U$ (towards Galactic centre), $V$ (in the direction of rotation) and $W$ (towards
the north Galactic pole).}
\end{table*}

\subsection{Mean velocity and the asymmetric drift}
\label{sec:4.1}
\begin{figure}
 \includegraphics[width=\columnwidth]{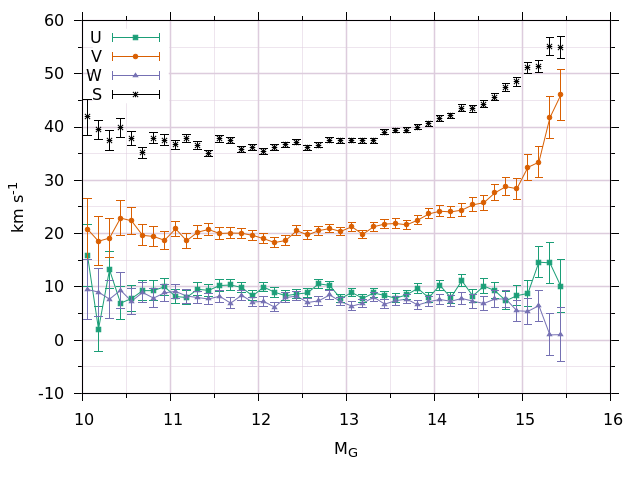}
 \caption{The $U,V,W$ components of the mean Solar motion with respect to the WDs, including a measure of the scalar velocity
 dispersion $S$. The trend in $V$ and $S$ is a manifestation of the asymmetric drift arising from the magnitude / age correlation.}
 \label{fig:mean_motion}
\end{figure}

In figure~\ref{fig:mean_motion} we plot the $U,V,W$ components of the mean Solar motion with respect to the WDs, along with
the quantity $S$ from DB98 that is a scalar measure of the velocity dispersion. The most obvious feature is the rise in the
$V$ component towards faint magnitudes, which is a manifestation of the asymmetric drift. This is accompanied by a corresponding
rise in the velocity dispersion as some of the original circular motion is scattered randomly into the $U,W$ components.
At magnitudes brighter than $M_G\approx13$ there is no trend in $V$ or $S$; in this range the correlation between magnitude
and mean age is nonexistent.
No break analogous to Paranego's discontinuity is present; there is no equivalent of the MS turnoff colour in the WD population.
The $U,W$ components show no systematic variation with $M_G$ as expected. Less obvious features
include the deviation in $U$ and $W$ beyond $M_G\approx15$; this is likely due to contamination by spheroid and thick disk WDs
that make up a larger fraction of the population at these magnitudes.

\subsection{Velocity dispersion and the local standard of rest}
\label{sec:4.2}
Classically, the main utility of measurements of the mean motion versus velocity dispersion is in estimating the Solar
motion with respect to the local standard of rest. This exploits the linear relationship between the 
velocity dispersion $S^2$ and rotational lag first identified empirically by \citet{stromberg1946}; the modern theoretical
framework for the asymmetric drift relation is described in \citet{bt2008} (see equation 4.228).
In practise this is done by extrapolating the mean velocity to $S^2=0$, which provides an estimate of the Solar motion
relative to a (hypothetical) population of stars with zero velocity dispersion. Such a population represents newly born
stars on closed, circular orbits, the motion of which defines the local standard of rest in the solar neighbourhood.

\begin{figure}
 \includegraphics[width=\columnwidth]{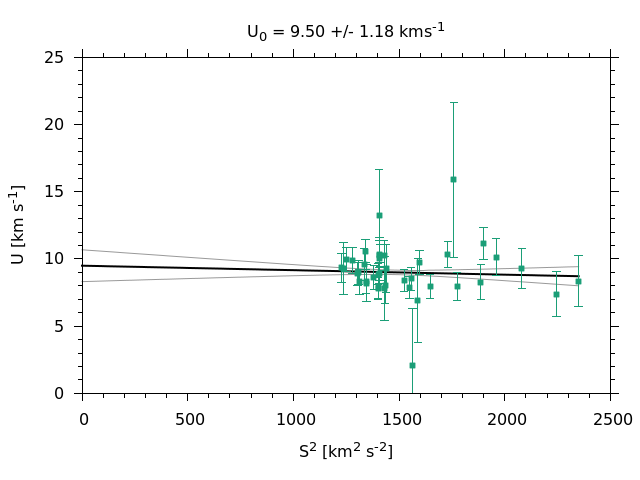}
 \caption{$U$ component of the mean Solar motion versus velocity dispersion; the black line is a linear fit and the
 grey lines indicate one-sigma confidence intervals.}
 \label{fig:lsr_u}
\end{figure}

\begin{figure}
 \includegraphics[width=\columnwidth]{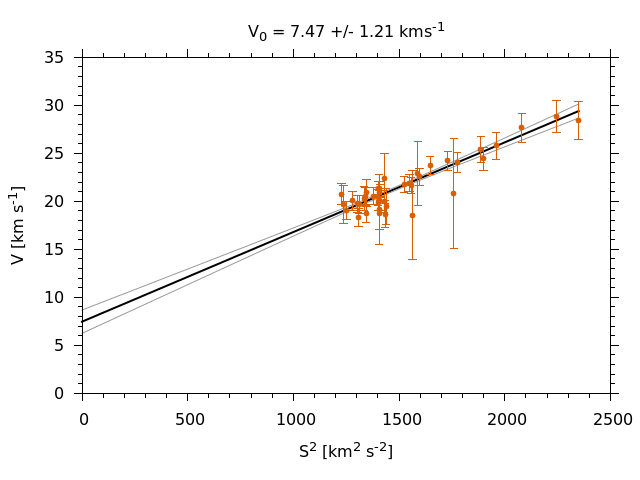}
 \caption{$V$ component of the mean Solar motion versus velocity dispersion.}
 \label{fig:lsr_v}
\end{figure}

\begin{figure}
 \includegraphics[width=\columnwidth]{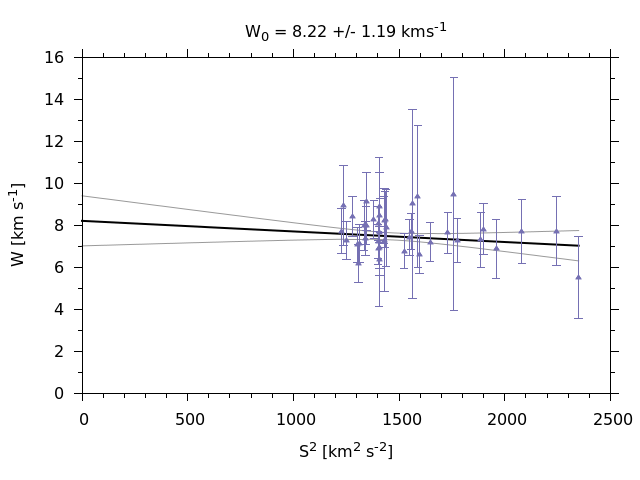}
 \caption{$W$ component of the mean Solar motion versus velocity dispersion.}
 \label{fig:lsr_w}
\end{figure}

Figures~\ref{fig:lsr_u}--~\ref{fig:lsr_w} plot the $U,V,W$ components of the mean Solar motion against $S^2$ along
with linear fits. In these plots we have omitted the four points fainter than $M_G=15.0$ for the reasons outlined in
section~\ref{sec:4.1}. For the $U$ and $W$ components there is no strong correlation and the fit is flat within one sigma,
whereas for $V$ there is a clear slope. Extrapolating to $S^2=0$ yields 
$(U,V,W)_{\odot} = (9.50\pm1.18, 7.47\pm1.21, 8.22\pm1.19)$ kms$^{-1}$ for the Solar motion with respect to the
local standard of rest.
Note that relative to similar studies that use MS stars we have few points at low $S^2$ due to the WDs having a certain 
minimum age, which results in larger uncertainties due to extrapolating over a longer interval.

\subsection{Velocity ellipsoid}
\label{sec:vel_ellip}
\begin{figure}
 \includegraphics[width=\columnwidth]{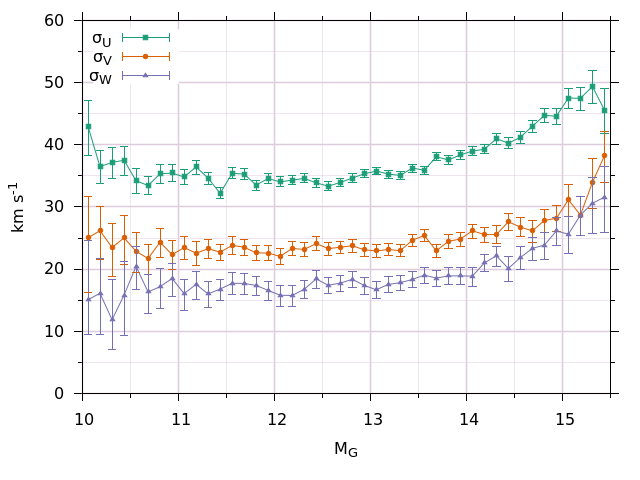}
 \caption{Components of the velocity dispersion tensor along the axes of the Galactic coordinate frame.}
 \label{fig:vel_disp_tens}
\end{figure}

In figure~\ref{fig:vel_disp_tens} we plot the diagonal components of the velocity dispersion tensor against magnitude.\footnote{In this figure and the ones that follow, the plotted quantities are nonlinear functions of our estimated values:
the (asymmetric) errorbars depict the $15.87$ and $84.13$ percentiles of the error distribution, obtained from Monte Carlo integration.}
In all components the velocity dispersion increases towards faint magnitudes, although it seems the rate of increase in the
$V$ component is smaller. The velocity dispersion is in general larger than that of MS stars, as the mean age of WDs is larger due
to the lack of young, newly formed stars: towards the faintest magnitudes nearly all the WDs are very old.

\begin{figure}
 \includegraphics[width=\columnwidth]{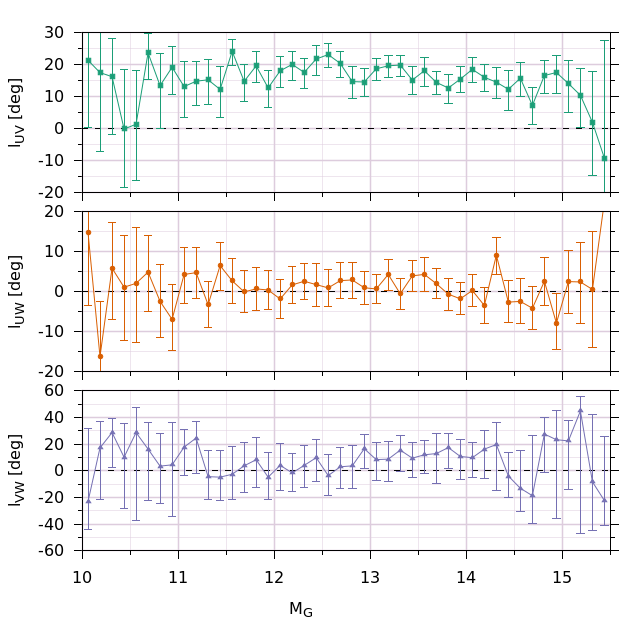}
 \caption{Vertex deviation in each of the  $UV$, $UW$ and $VW$ planes. The deviations in the $UW$ and $VW$ planes, which involves the velocity 
 dispersion in the direction out of the plane, are consistent with zero, whereas a nonzero deviation of around 15 degrees is observed in the 
 $UV$ plane.}
 \label{fig:vert_dev}
\end{figure}

\begin{figure}
 \includegraphics[width=\columnwidth]{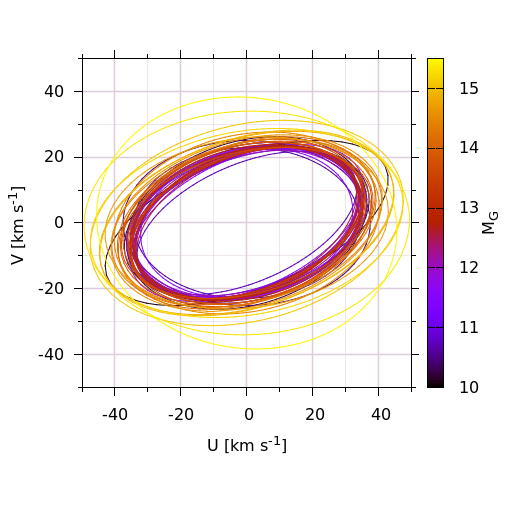}
 \caption{Projection of the velocity ellipsoid onto the $UV$ plane. Ellipses are drawn at the one-sigma level.}
 \label{fig:proj_uv}
\end{figure}

\begin{figure}
 \includegraphics[width=\columnwidth]{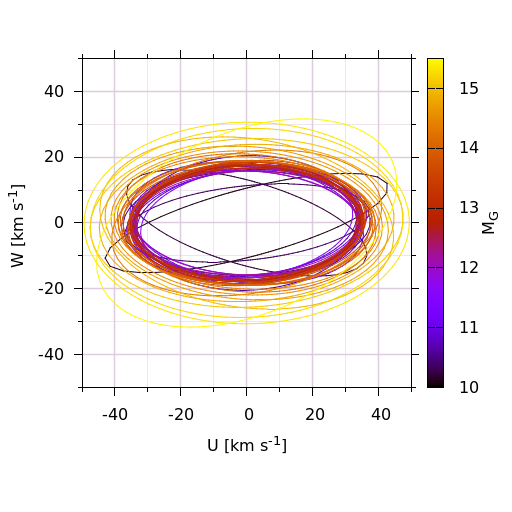}
 \caption{Projection of the velocity ellipsoid onto the $UW$ plane, as per figure~\ref{fig:proj_uv}}
 \label{fig:proj_uw}
\end{figure}

\begin{figure}
 \includegraphics[width=\columnwidth]{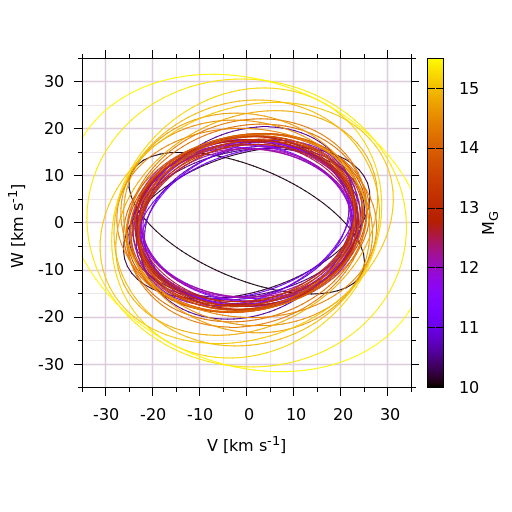}
 \caption{Projection of the velocity ellipsoid onto the $VW$ plane, as per figure~\ref{fig:proj_uv}}
 \label{fig:proj_vw}
\end{figure}

The velocity dispersion tensor has some significant off-diagonal elements, indicating that the velocity ellipsoid is not aligned
with the $U,V,W$ axes of the Galactic frame. This can be quantified by the \textit{vertex deviation} $l_{XY}$, which is the angle
by which one has to rotate the velocity ellipsoid in the $XY$ plane to bring it into alignment. The vertex deviation in each of
the $UV$, $UW$ and $VW$ planes is plotted in figure~\ref{fig:vert_dev}. The deviations in the $UW$ and $VW$ planes, which involve the velocity 
dispersion in the direction out of the plane, are consistent with zero, whereas a nonzero deviation of around 15$^{\circ}$ is observed in the 
$UV$ plane. These results are similar to those of DB98,
who find $l_{UV}\approx20^{\circ}$ for early spectral types and $10^{\circ} \pm 4^{\circ}$ for old disk stars,
although we find that the vertex deviation in $UV$ is roughly constant with
no obvious trend with magnitude and, equivalently, mean age. Figures~\ref{fig:proj_uv}--~\ref{fig:proj_vw} present projections of the
velocity dispersion tensor into each of the $UV$, $UW$ and $VW$ planes, which helps to visualise the velocity ellipsoid orientation, scale
and evolution with magnitude. Ellipses are drawn at the one-sigma level.

\section{Discussion}
\label{sec:disc}
Our values for $U_{\odot}$ and $W_{\odot}$ agree within one sigma with the values obtained by both \citet{schonrich2010} and DB98.
However, our value for $V_{\odot}=7.47\pm1.21$ differs from the \citet{schonrich2010} value of $V_{\odot}=12.24\pm0.47$ by $\approx3.7\sigma$
and from the \citet{kawata2019} value of $V_{\odot}=12.4\pm0.7$ by a similar amount,
being more in line with the value of $V_{\odot}=5.25\pm0.62$ obtained by DB98. As explained in \citet{schonrich2010} the DB98 
value for $V_{\odot}$ is affected by systematics arising from the metallicity gradient in the disc; briefly, metal-rich stars
generally formed at $R<R_0$ and visit the Solar neighbourhood close to the apocentre of their orbit where their azimuthal
velocities lag behind the local circular speed. Increasing metallicity shifts the MS to the right in the colour-magnitude plane,
introducing a velocity gradient across the MS and biasing the color/age correlation.
The fact that our value of $V_{\odot}$ lies between the DB98 and \citet{schonrich2010} value could indicate that our WD sample
is affected to a lesser extent by similar systematics. Although the location of a WD in the colour-magnitude diagram is independent of
the progenitor metallicity to first order, in a population of stars the metallicity-dependent progenitor lifetimes
could play a role in introducing kinematic substructures in the WD colour-magnitude plane similar to those described by
\citet{schonrich2010}.
In addition, over the magnitude range of interest to this study the Gaia WD sequence splits into two branches, due mostly to atmospheric
composition but also to the presence of massive WDs possibly formed through stellar mergers \citep{kilic2018, elbadry2018}. The question 
of to what extent this affects the kinematics as a function of $M_G$ is complex. The WD cooling rate depends on both atmosphere
composition and mass, so both of these properties will affect the age distribution of stars as a function of $M_G$. While the
the kinematics are likely to be independent of atmosphere type (DA / non-DA), massive WDs formed from merged binary systems
could conceivably have different kinematics. Ultimately, both of these effects could lead to kinematic substructure in the WD
colour-magnitude diagram and be responsible for systematic errors in our results.
However, a thorough investigation of this is outwith the scope of our study.
%
% Response to reviewer point 5)
%\textbf{
It is also possible that large scale streaming motions within the disk are affecting our results, due to the local nature of
our sample. For example, \citet{bovy2015} finds that the Solar neighbourhood is likely affected by a streaming motion of the
order of 10kms$^{-1}$ in the direction of rotation, which would bias our estimate of $V_{\odot}$.
%}

The nonzero vertex deviation in the $UV$ plane is understood to be caused by the non-axisymmetric Galactic potential in the
disc, likely due to the effect of the spiral arms. The fact that the vertex deviation is independent of magnitude indicates
that this has been stable for a long period. No such asymmetry in the potential in the $UW$ and $VW$ planes is evident.

Previous WD proper motion surveys and luminosity function studies that used RPM selection generally assume a constant value
for the reflex Solar motion and WD velocity distribution when computing correction factors for stars that fall
below the $v_{\mathrm{tan}}\approx30$kms$^{-1}$ selection threshold. This includes \citealt{harris2006}; RH11; \citealt{lam2019}.
As shown in this study, there is a strong correlation between WD absolute magnitude and both the mean and dispersion of the
heliocentric velocity, which will cause the correction factors to be somewhat overestimated towards the faint end of the
luminosity function, leading to an overestimation of the spatial density for faint WDs.

For example, the study of RH11 adopted a mean velocity of $(U,V,W)_{\odot} = (8.62, 20.04, 7.1)$ kms$^{-1}$ and velocity
dispersion $(\sigma_U, \sigma_V, \sigma_W) = (32.4, 23.0, 18.1)$ kms$^{-1}$ with no off-diagonal terms. In conjunction with their
lower tangential velocity threshold of $30$kms$^{-1}$ this leads to a sky-averaged discovery fraction of $\approx57$\%, which
is then used to correct the estimated spatial density to account for the missing low velocity stars.
% df = 0.43166598645001525
% df = 0.31755628207444575
However for faint WDs both the mean velocity and velocity dispersion are significantly larger than these values; adopting instead
the values that we measure for the $M_G$ bin centred on $14.563$ we get a sky-averaged discovery fraction of $\approx68$\%, meaning that
the RH11 study will to first order overestimate the spatial density of WDs of these magnitudes by around a factor $\approx1.2$,
although the true figure depends on the complex effects of the magnitude-dependent proper motion limits when computing the
generalised survey volume, as described in \citet{lam2015}.

\section{Conclusions}
\label{sec:conc}
We have presented the first detailed analysis of the kinematics of Galactic disc WD stars in the Solar neighbourhood,
based on a large, kinematically unbiased sample of WDs with absolute astrometry provided by Gaia DR2.
Various classical properties of the Solar neighbourhood kinematics have been detected for the first time
in the WD population.
These are found to be largely consistent with expectations based on MS star studies, although form an independent measure.
The results of this study provide an important input to proper motion surveys for white dwarfs, which require knowledge
of the velocity distribution in order to correct for missing low velocity stars that are culled from the sample to
reduce subdwarf contamination.

\section*{Acknowledgements}

This work has made use of data from the European Space Agency (ESA) mission
{\it Gaia} (\mbox{\url{https://www.cosmos.esa.int/gaia}}), processed by the {\it Gaia}
Data Processing and Analysis Consortium (DPAC,
\url{https://www.cosmos.esa.int/web/gaia/dpac/consortium}). Funding for the DPAC
has been provided by national institutions, in particular the institutions
participating in the {\it Gaia} Multilateral Agreement.

%%%%%%%%%%%%%%%%%%%%%%%%%%%%%%%%%%%%%%%%%%%%%%%%%%



\bibliographystyle{mnras}
\bibliography{main} % if your bibtex file is called example.bib

%%%%%%%%%%%%%%%%%%%%%%%%%%%%%%%%%%%%%%%%%%%%%%%%%%

%%%%%%%%%%%%%%%%% APPENDICES %%%%%%%%%%%%%%%%%%%%%

\appendix
\section{ADQL query}
\label{app:adql}
The ADQL query used to select our initial sample from the Gaia archive is shown below. This returns a table containing $3061480$ rows,
which is subject to additional filtering as described in section~\ref{sec:wd_selection}.

\begin{verbatim}
SELECT *
FROM   gaiadr2.gaia_source
WHERE parallax_over_error > 5.0
 AND phot_bp_mean_flux_over_error > 10
 AND phot_rp_mean_flux_over_error > 10
 AND astrometric_n_good_obs_al > 5
 AND phot_bp_rp_excess_factor BETWEEN 
     1.0 + (0.03*POWER(bp_rp, 2.0))
     AND 1.3 + (0.06*POWER(bp_rp, 2.0))
 AND 1000.0/parallax < 250
\end{verbatim}

%%%%%%%%%%%%%%%%%%%%%%%%%%%%%%%%%%%%%%%%%%%%%%%%%%

% Don't change these lines
\bsp	% typesetting comment
\label{lastpage}
\end{document}